\documentclass[prb,preprint,eqsecnum]{revtex4}%
\usepackage{amsmath}
\usepackage{graphicx}
\usepackage{amssymb}
\usepackage{amsfonts}%
\begin{document}
\title{Coulomb potentials in two and three dimensions under periodic boundary conditions}
\author{Sandeep Tyagi}
\email{satst27@pitt.edu}
\affiliation{Department of Physics and Astronomy, University of Pittsburgh, Pittsburgh,
Pennsylvania 15260}

\begin{abstract}
A method to sum over logarithmic potential in 2D and Coulomb potential in 3D
with periodic boundary conditions in all directions is given. We consider the
most general form of unit cells, the rhombic cell in 2D and the triclinic cell
in 3D. For the 3D case, this paper presents a generalization of Sperb's work
[R. Sperb, Mol. Simulation, \textbf{22}, 199-212(1999)]. The expressions
derived in this work converge extremely fast in all region of the simulation
cell. We also obtain results for slab geometry. Furthermore, self-energies for
both 2D as well as 3D cases are derived. Our general formulas can be employed
to obtain Madelung constants for periodic structures.

\end{abstract}
\maketitle

\section{Introduction}

It has become a common practice to employ numerical simulations in the study
of physical problems, which are difficult to solve analytically. Since it is
not possible to simulate realistic physical systems, containing ions of the
order of Avogadro number, one usually works with a very small system. For
small systems, containing a few hundred to a few thousand charges, boundary
effects become relatively pronounced, especially if the nature of interaction
is long range. To avoid this problem, periodic boundary conditions (PBC) are
usually employed. In many simulations, the nature of interaction is such that
the potential satisfies the Poisson equation. For example, a logarithmic
interaction in two dimensions (2D) and a Coulomb potential in three dimensions
(3D) both satisfy the Poisson equation in 2D and 3D respectively. We refer to
a potential which goes as $r^{\left(  -d+2\right)  }$ in a $d\geq2$
dimensional isotropic space as a Coulomb type potential. The Coulomb type
potentials fall under the category of long range potentials. In fact, in a
$d\geq2$ dimensional space, any interaction which goes as $r^{-\alpha},$ where
$\alpha<\left(  d-1\right)  $ is known as a long range interaction. The reason
being, while the potential decays as $r^{-\alpha}$, the volume element goes as
$r^{\left(  d-1\right)  }$. As a result, in a periodic system, even charges
located at infinity give rise to a finite contribution to energy and forces,
which cannot be neglected. We consider Coulomb type of potentials in 2D and 3D
in this paper.

To derive a formula for interaction between two particles with PBC imposed,
one has to consider the interaction of a particle with periodic repetitions of
itself, as well as that of the second particle. Interaction energy of a
particle with its own periodic repetitions is termed as the self-energy.
Determination of self-energy is important in simulations where the size of
simulation box may change during the simulation. For example, such a case
arises in an isobaric Monte Carlo simulation. The aim of this paper is to
consider the kind of interactions mentioned above for the most general type of
unit cells in 2D and 3D. We consider a rhombic unit cell in 2D and a triclinic
cell in 3D, with origin lying at the bottom left corner of the unit cell. The
unit cell contains a number of ions, which interact via Coulomb type
potential, satisfying the Poisson equation in their respective dimensions. The
unit cell repeats itself in all directions under PBC. Hence, the interaction
of a particle located at $\mathbf{r}$ with another particle located at the
origin includes, apart from the direct interaction between the two particles,
the interaction of the first particle with all periodic images of the second
particle. These periodic images of the second particle are located at lattice
vector sites given by $\mathbf{l}=m\mathbf{a}+n\mathbf{b}+p\mathbf{c}$, where
$m$, $n$ and $p$ range from $-\infty$ to $+\infty$. Also, the particle
interacts with its own images located at $\mathbf{r}+\mathbf{l}$, where
$\mathbf{l}$ is defined as above. Thus, if we have $N$ charges $q_{i}$ in a
charge neutral unit cell, then the Coulomb energy may be written as%
\begin{equation}
E=\frac{1}{2}\sum_{\mathbf{n}\in%
\mathbb{Z}
^{3}}^{\prime}\sum_{i,j=1}^{N}\frac{q_{i}q_{j}}{\left\vert \mathbf{r}%
_{i}-\mathbf{r}_{j}+\mathbf{n}\right\vert },\label{first}%
\end{equation}
\qquad where a prime indicates that $\mathbf{n}$ $=0$ term is to be excluded
for the case when $i=j$. The series in Eq.(\ref{first}) is a conditional
series. This series can be summed up to any value depending on the order in
which the terms of the series are grouped. Therefore, a summation convention
has to be specified based on the physical nature of the problem in mind.

The conditional series mentioned above may be evaluated by introducing
background charges in a way that the total background charge adds up to zero.
Imposing background charges in this way leads to well defined ways of summing
the conditional series. However, results of the summation of conditional
series may still differ in view of the method employed to impose background
charges, as the background charges may have a structure of their own. For
example, background charges in a 3D system with PBC may be imposed in the
following two ways. A charge $q$ and all its periodic repetitions under the
PBC may be viewed as a set of layers along an axis of the unit cell. In order
to impose background charge on this system, we may assume that all these
different layers are charge neutral separately. Thus, we may assume that for a
layer composed of charge $q$ and its periodic images, one has an additional
uniform charge density of $-q/a,$ where $a$ is the area of the 2D unit cell.
Thus the overall charge contained in each 2D cell of the layer is zero.
Another charge $q^{\prime}$ present in the system will interact with the set
of charges $q$ as well as the neutralizing background surface charge with
charge density. However, it can be shown that introducing these uniform
background charge sheets leads to some unwanted terms, as the sheets have a
structure of their own.

However, there is a better way of imposing background charges, without
introducing any structure of the background charges themselves. This can be
achieved by distributing a uniform 3D charge on the grid made out of charge
$q$ and its images. The neutralizing background charge now has a uniform
charge density of $-q/V$, where $V$ is the volume of the unit cell. This
volume charge adds up to zero at any point due to the overall charge
neutrality condition and thus does not introduce any artificial structure,
such as the uniform sheets in the previous case.

The results of the two prescriptions suggested above differ by two terms. The
first term depends on the square of the component of the dipole
moment\cite{smith}, along the direction of layering, of the original charges
contained in the unit cell. The second term depends linearly on the distance
between the pair of charges along the direction of layering.

In this paper, we adopt the second procedure. Using the results derived here,
it will be easy to establish connection between the results of two summation
conventions mentioned above. Introduction of neutralizing charge background in
the form of a uniform cloud leads to only the intrinsic part\cite{lekner1} of
potential energy and this technique has previously been employed by
Lekner\cite{lekner1} and Sperb\cite{sperb2}. It is important to know that the
two procedures mentioned above still do not lead to the correct energy of a
collection of charges interacting under the PBC, if one wants a limit of
spherical means \cite{sperb2}. De Leeuw \textit{et. al}\cite{deleeuw} have
shown that for 3D case, an extra term depending on the total dipole moment of
the unit cell has to be added to get the correct energy of
charges. For the 2D case the correction term turns out to be zero.

With the help of discussion above, the energy of $N$ particles contained in a
unit cell with periodic boundaries and interacting through a Coulomb type
potential in 3D can be expressed as,
\begin{equation}
E_{\text{total}}=\frac{1}{2}\sum_{i,j;i\neq j}q_{i}q_{j}G(\mathbf{r}%
_{i}-\mathbf{r}_{j})+\sum_{i}q_{i}^{2}G_{\text{self}}+\frac{2\pi}{3}\left(
\sum_{i}q_{i}\mathbf{r}_{i}\right)  ^{2}.\label{tot}%
\end{equation}
Here the charges are denoted by $q_{i}$'s and their positions in the unit
cell by $\mathbf{r}_{i}$'s and $1\leq i\leq N$. The last term in
Eq.(\ref{tot}) is the dipole term introduced by De Leeuw \textit{et.
al}\cite{deleeuw} . For the 2D case one has only the first two terms on the
right hand side. Our aim in this paper is to obtain expressions for
$G(\mathbf{r})$ and $G_{\text{self}}$ in 2D and 3D.

Before proceeding further, we briefly discuss three main approaches in use to
obtain Coulomb interaction with periodic boundaries. These three approaches
are due to Ewald\cite{ewald}, Lekner \cite{lekner1,lekner2} and Sperb\cite{sperb2}.
The Ewald method was developed eighty years ago in connection with the
evaluation of Madelung constants. This method, in spite of its shortcomings,
is still very much in use. The method proceeds by breaking the original
summation in two parts. One of these sums is carried out in real space and
another one in Fourier space. This splitting of summation depends on a real
parameter which has to be chosen judiciously, failing which the series in real
and Fourier space might converge very slowly. In general, to calculate a
pair-wise interaction it usually requires a few hundred terms involving
complementary error functions.

An alternative to the Ewald method was given by Lekner\cite{lekner1}. This
method involves an evaluation of a few dozen terms if the position vector
$\mathbf{r}$ is not very small. If $\mathbf{r}$ tends to zero this method
converges slowly. The problem of convergence was fixed later by Lekner in
another paper\cite{lekner2}, following Sperb's work\cite{sperb1}. Lekner
method has not been generalized to a triclinic cell yet. Though, it is
possible to generalize Lekner's work for a triclinic cell, here, we take a
different approach along the lines of Ref.\onlinecite{paper3} to obtain
results for a triclinic cell.

Among the latest advances on Coulomb sums is by Gr{\o }nbech-Jensen
\cite{niels} in 2D and Sperb\cite{sperb2} in 3D. Sperb's results are similar
to that of Harris \textit{et al.}\cite{harris} and Crandall \textit{et al.}
\cite{crandall}. A major advantage of Sperb's work is that it can be employed
to get $N\ln(N)$ scaling in time\cite{strebel}, where $N$ is the number of
ions present in the system. On the other hand, with Ewald summation method,
one can get only $[N\ln(N)]^{3/2}$ scaling\cite{ceperley}. In this paper, our aim is to generalize
Sperb's work\cite{sperb2} to a triclinic cell. The method given in this work
will contain Sperb's result in a simplified form as a special case. Also for
the first time, an alternative to Ewald's technique will be given, which can
be applied to the most general kind of unit cell in a computer simulation, a
triclinic unit cell. We will also discuss scaling of $N\ln(N)$ that may be
achieved with the use of formulas developed here.

\begin{figure}[tbh]
\includegraphics[angle=-90,scale=0.80]{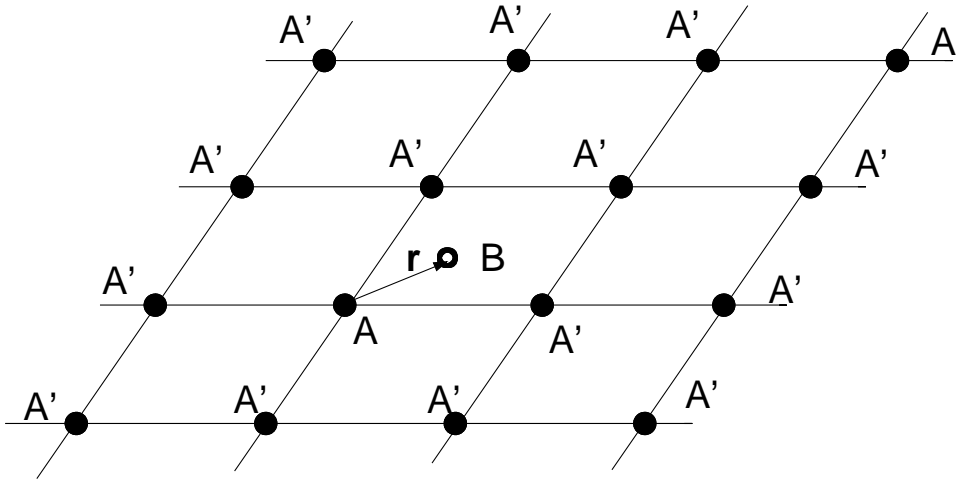}\caption{A schematic
diagram explaining the set of for a 2D rhombic cell. A charge located at B
interacts with another charge located at the origin A as well as its periodic
images located at A's.}%
\label{layers}%
\end{figure}

\section{Logarithmic interaction in 2D}

An excellent method to sum over Coulomb type potential (logarithmic
interaction) in 2D was given by Gr{\o }nbech-Jensen\cite{niels}. Another
alternative was provided by Tyagi \textit{et al.} \cite{paper2} in a recent
paper. The problem with the later approach is that the lattice sum does not
converge when the two charges are close together within the unit cell. This
problem will be addressed here and formulas will be modified in such a way
that the convergence is achieved for even those cases where charges are close
to each other. Thus we will obtain a result which is different from
Gr{\o }nbech-Jensen but still as efficient.

We consider a rhombic cell with periodic boundaries along the $x$ and $y$
directions. A sketch of the cell is shown in Fig. 1. A particle, located at
position $\mathbf{r,}$ interacts logarithmically with charges located at the
vertices of a rhombic grid. A formula was developed in Ref.\onlinecite{paper2}
to compute this sum. We sketch a portion of that derivation here for the sake
of completeness. Consider the Poisson equation in 2D:%

\begin{equation}
\nabla^{2}G(\mathbf{r})=-2\pi\sum_{l}\delta(\mathbf{r}+\mathbf{l})+\frac{2\pi
}{l_{1}l_{2}\sin\theta}.\label{e01}%
\end{equation}
The second term on the right hand side amounts to the presence of a
neutralizing background charge. The solution of Eq. (\ref{e01}) is given by
\begin{equation}
G(\mathbf{r})=\frac{2\pi}{l_{1}l_{2}\sin\theta}\lim_{\xi\rightarrow0}\left(
\sum_{\mathbf{Q}}\frac{\exp(i\mathbf{Q}\cdot\mathbf{r})}{\mathbf{Q}^{2}%
+\xi^{2}}-\frac{1}{\xi^{2}}\right)  ,\label{expansion}%
\end{equation}
where $l_{1}$ and $l_{2}$ denote the lengths of the sides of the rhombic cell
and
\begin{equation}
\mathbf{r}=r_{1}\mathbf{e}_{1}+r_{2}\mathbf{e}_{2},\,\,\,\,\,\mathbf{Q}%
=n_{1}\mathbf{b}_{1}+n_{2}\mathbf{b}_{2}.\label{e02}%
\end{equation}
Here, $0\leq r_{1}<l_{1}$, $0\leq r_{2}<l_{2}$ and $\mathbf{e}_{1}$ and
$\mathbf{e}_{2}$ represent the unit vectors along the axis of the rhombic
cell, with $\mathbf{e}_{1}.\mathbf{e}_{2}=\cos\theta$. We have also introduced
an infinitesimal parameter $\xi$. The sum over $\mathbf{Q}$ runs over all
reciprocal lattice vectors spanned by
\begin{equation}
\mathbf{b}_{i}=\frac{2\pi}{l_{i}\sin^{2}\theta}(\mathbf{e}_{i}-\mathbf{e}%
_{j}\cos\theta),\label{e03}%
\end{equation}
for $(i,j)=(1,2),(2,1)$ and $n_{1}$ and $n_{2}$ are integers. Introduction of
an infinitesimal parameter $\xi$ as in Eq. (\ref{expansion}) implies
assumption of the presence of a neutralizing background charge. Thus, here a
charge $q$ located at $\left(  x,y\right)  $ in the unit rhombic cell
interacts with charges $q^{\prime}$ located at the origin and all other
vertices of the grid. The charge $q$ also interacts with a uniform layer of
background charge superimposed on the grid of $q^{\prime}$ charges such that
the charge density is $-q^{\prime}/a,$ where $a=l_{1}l_{2}\sin\theta,$ is the
area of the unit cell. From now onward we will always assume that a final
limit $\xi\rightarrow0$ is to be taken. Using the value $\mathbf{Q}$ from
Eq.(\ref{e02}) and Eq.(\ref{e03}) in Eq.(\ref{expansion}) we obtain
\begin{equation}
G(\mathbf{r})=\frac{\sin\theta}{2\pi l_{1}l_{2}}\sum_{n_{1}=-\infty}^{\infty
}\sum_{n_{2}=-\infty}^{\infty}\frac{\exp\left[  i2\pi\left(  n_{1}\frac{r_{1}%
}{l_{1}}+n_{2}\frac{r_{2}}{l_{2}}\right)  \right]  }{\left(  \frac{n_{1}%
}{l_{1}}\right)  ^{2}-2\frac{n_{1}}{l_{1}}\frac{n_{2}}{l_{2}}\cos
\theta+\left(  \frac{n_{2}}{l_{2}}\right)  ^{2}+\frac{\xi^{2}}{l_{1}l_{1}}%
}-\frac{\sin\theta}{2\pi\sigma}\frac{1}{\xi^{2}},\nonumber
\end{equation}
where $0\leq r_{i}/l_{i}<1$, $\sigma=l_{2}/l_{1}$ and we have redefined
infinitesimal parameter $\xi$ for sake of calculations. We now evaluate the
sum
\begin{align}
f(n_{1},\xi) &  =\sum_{n_{2}=-\infty}^{\infty}\frac{\exp\left(  i\frac{2\pi
}{l_{2}}n_{2}r_{2}\right)  }{\left(  \frac{n_{1}}{l_{1}}\right)  ^{2}%
-2\frac{n_{1}}{l_{1}}\frac{n_{2}}{l_{2}}\cos\theta+\left(  \frac{n_{2}}{l_{2}%
}\right)  ^{2}+\frac{\sigma^{2}\xi^{2}}{l_{2}^{2}}}\label{sum1}\\
&  =l_{2}^{2}\sum_{n_{2}=-\infty}^{\infty}\frac{\exp\left(  i\frac{2\pi}%
{l_{2}}n_{2}r_{2}\right)  }{n_{1}^{2}\sigma^{2}-2n_{1}n_{2}\sigma\cos
\theta+n_{2}^{2}+\sigma^{2}\xi^{2}}\nonumber\\
&  =\pi l_{2}^{2}\exp(i2\pi\beta_{n_{1}}t_{2})\frac{\exp(-i2\pi\beta_{n_{1}%
})\sinh\left[  \gamma_{n_{1}}t_{2}\right]  +\sinh\left[  2\pi\gamma_{n_{1}%
}(1-t_{2})\right]  }{\gamma_{n_{1}}\left[  \cosh(2\pi\gamma_{n_{1}})-\cos
(2\pi\beta_{n_{1}})\right]  },\nonumber
\end{align}
where $t_{2}=r_{2}/l_{2}$,%

\begin{equation}
\,\beta_{n}=n\sigma\cos\theta,\,\,\gamma_{n}=\sigma\sqrt{(n^{2}\sin^{2}%
\theta+\xi^{2})}, \label{e05}%
\end{equation}
and we have used the identity (here $\alpha<2\pi$)
\begin{equation}
\sum_{n=-\infty}^{\infty}\frac{\exp\left(  in\alpha\right)  }{\left(
n-\beta\right)  ^{2}+\gamma^{2}}=\frac{\pi}{\gamma}\frac{\exp\left[
i\beta\left(  \alpha-2\pi\right)  \right]  \sinh\left(  \gamma\alpha\right)
+\exp\left(  i\beta\alpha\right)  \sinh\left[  \gamma\left(  2\pi
-\alpha\right)  \right]  }{\cosh\left(  2\pi\gamma\right)  -\cos\left(
2\pi\beta\right)  },%
\end{equation}
which is derived in Appendix A. The sum defined in Eq.(\ref{sum1}) can be
written as%

\begin{align}
G(\mathbf{r})  &  =\frac{\sigma\sin\theta}{2}\sum_{n=-\infty}^{+\infty}%
\exp\left[  i2\pi\left(  nt_{1}+\beta_{n}t_{2}\right)  \right] \label{e06}\\
&  \times\frac{\exp(-2\pi\beta_{n})\sinh(2\pi\gamma_{n}t_{2})+\sinh\left[
2\pi\gamma_{n}(1-t_{2})\right]  }{\gamma_{n}\left[  \cosh(2\pi\gamma_{n}%
)-\cos(2\pi\beta_{n})\right]  }-\frac{\sin\theta}{2\pi\sigma}\frac{1}{\xi^{2}%
},\nonumber
\end{align}
where $t_{1}=r_{1}/l_{1}$. Separating out the term corresponding to $n=0$, we obtain%

\begin{align}
G(\mathbf{r}) &  =\frac{\sigma\sin\theta}{2}\left(  \frac{\sinh(2\pi\xi
t_{2})+\sinh\left[  2\pi\xi(1-t_{2})\right]  }{\xi\left[  \cosh(2\pi
\xi)-1\right]  }\right)  -\frac{\sin\theta}{2\pi\sigma}\frac{1}{\xi^{2}%
}\label{e07}\\
&  +\frac{\sigma\sin\theta}{2}\sum_{n}^{\prime}\exp\left(  i2\pi nt\right)
\frac{\exp(-2\pi\beta_{n})\sinh(2\pi\gamma_{n}t_{2})+\sinh\left[  2\pi
\gamma_{n}(1-t_{2})\right]  }{\gamma_{n}\left[  \cosh(2\pi\gamma_{n}%
)-\cos(2\pi\beta_{n})\right]  },\nonumber
\end{align}
where $t=t_{1}+t_{2}\sigma\cos\theta$ and a prime on the summation sign
indicates that the term corresponding to $n=0$ is not to be included. Taking
the limit $\xi\rightarrow0,$ one obtains%

\begin{equation}
\lim_{\xi\rightarrow0}\left(  \frac{\sinh\left[  2\pi\xi\sigma t_{2}\right]
+\sinh\left[  2\pi\xi\sigma(1-t_{2})\right]  }{\xi\sigma\left[  \cosh(2\pi
\xi\sigma)-1\right]  }-\frac{1}{\pi\sigma^{2}\xi^{2}}\right)  =\frac{\pi}%
{3}\left(  1-6t_{2}+6t_{2}^{2}\right)  .
\end{equation}
\qquad Thus we obtain the following expression for $G(\mathbf{r})$%

\begin{align}
G(\mathbf{r}) &  =\frac{\sigma\sin\theta}{2}\frac{\pi}{3}\left(
1-6t_{2}+6t_{2}^{2}\right)  \label{g1}\\
&  +\frac{\sigma\sin\theta}{2}\sum_{n}^{\prime}\exp\left(  i2\pi nt\right)
\frac{\exp\left(  -i2\pi\beta_{n}\right)  \sinh\left(  2\pi\gamma_{n0}%
t_{2}\right)  +\sinh\left[  2\pi\gamma_{n0}(1-t_{2})\right]  }{\gamma
_{n0}\left[  \cosh(2\pi\gamma_{n0})-\cos(2\pi\beta_{n})\right]  },\nonumber
\end{align}
where $\gamma_{n0}=\sigma\left\vert n\sin\theta\right\vert $. Due to the
symmetrical nature of the unit cell, it suffices to look at only that part of
the unit cell, which corresponds to $0\leq t_{1}\leq0.5$ and $0\leq t_{2}%
\leq0.5$. Eq.\ (\ref{g1}) fails to converge fast enough when $t_{2}%
\rightarrow0$. This problem can be easily fixed as follows. We add and
subtract the following term from Eq.\ (\ref{g1})%

\begin{equation}
h\left(  t,t_{2}\right)  =\frac{1}{2}\sum_{n}^{\prime}\frac{\exp(-2\pi
\gamma_{n0}t_{2})\exp(2\pi i n t)}{\left\vert n\right\vert }. \label{h01}%
\end{equation}
The quantity $h\left(  t,t_{2}\right)  $ can be easily evaluated by carrying
out the sum in Eq. (\ref{h01}) analytically. Using the identity%

\begin{equation}
\sum_{n=1}^{+\infty}\frac{\exp(-n \left\vert a\right\vert )\cos(n b)}%
{n}=-\frac{1}{2}\ln\left[  \cosh\left(  a\right)  -\cos\left(  b\right)
\right]  -\frac{\ln\left(  2\right)  }{2}+\frac{\left\vert a\right\vert }{2},
\label{h03}%
\end{equation}
one obtains%

\begin{align}
h\left(  t,t_{2}\right)   &  =-\frac{1}{2}\ln\left(  \cosh[2\pi t_{2}%
\sigma\sin\theta]-\cos[2\pi t]\right) \label{h04}\\
&  -\frac{\ln\left(  2\right)  }{2}+\frac{2\pi\sigma t_{2}\sin\theta}%
{2}.\nonumber
\end{align}
Thus we can write%

\begin{align}
G(\mathbf{r})=  &  \frac{\sigma\sin\theta}{2}\frac{\pi}{3}\left(
1-6t_{2}+6t_{2}^{2}\right) \label{h05}\\
&  -\frac{1}{2}\ln\left(  \cosh[\sigma\sin\left(  \theta\right)  2\pi
t_{2}]-\cos[2\pi t]\right)  +\frac{2\pi\sigma t_{2}\sin\theta}{2}-\frac
{\ln\left(  2\right)  }{2}\nonumber\\
&  \times\frac{1}{2}\sum_{n}^{\prime}\exp\left(  i2\pi nt\right)  \left\{
\frac{\exp(-i2\pi\beta_{n})\sinh(2\pi\gamma_{n0}t_{2})+\sinh\left[  2\pi
\gamma_{n0}(1-t_{2})\right]  }{\left\vert n\right\vert \left[  \cosh
(2\pi\gamma_{n0})-\cos(2\pi\beta_{n})\right]  }\right. \nonumber\\
&  \left.  -\frac{\exp(-2\pi\gamma_{n0}t_{2})}{\left\vert n\right\vert
}\right\}  ,\nonumber
\end{align}
After some effort, the above equation can be written as
\begin{align}
&  G(\mathbf{r})=\frac{\sigma\sin\theta}{2}\frac{\pi}{3}\left(  1+6t_{2}%
^{2}\right)  -\frac{1}{2}\ln\left(  \cosh\left[  2\pi\sigma\sin\left(
\theta\right)  t_{2}\right]  -\cos\left[  2\pi t\right]  \right) \label{h55}\\
&  -\frac{\ln\left(  2\right)  }{2}+\sum_{n=1}^{+\infty}\left\{  \cos\left[
2\pi\left(  n t-\beta_{n}\right)  \right]  \sinh(2\pi\gamma_{n0}t_{2}%
)+\cos(2\pi n t)\right. \nonumber\\
&  \left.  \times\left[  \exp\left(  -2\pi\gamma_{n0}t_{2}\right)  \cos\left(
2\pi\beta_{n}\right)  -\exp\left(  -2\pi\gamma_{n0}\right)  \cosh\left(
2\pi\gamma_{n0}t_{2}\right)  \right]  \right\}  /\nonumber\\
&  \left\{  n\left[  \cosh(2\pi\gamma_{n0})-\cos(2\pi\beta_{n})\right]
\right\}  .\nonumber
\end{align}
Equation (\ref{h55}) converges extremely fast for all values of $0\leq
t_{2}\leq0.5$. However, to achieve better convergence the sides of the rhombic
cell should be labelled such that $\sigma=l_{2}/l_{1}\geq1$. Now, an
expression for the self-energy can be easily obtained by taking the limits
$t_{1}\rightarrow0$ and $t_{2}\rightarrow0$ and subtracting
\begin{equation}
g\left(  \mathbf{r}\right)  =-\frac{1}{2}\ln\left(  r_{1}^{2}+r_{2}^{2}%
+2r_{1}r_{2}\cos\theta\right)  , \label{h06}%
\end{equation}
one obtains
\begin{align}
G_{\text{2d}}^{\text{self}}  &  =\lim_{\mathbf{r}\rightarrow0}\left(
G(\mathbf{r})-g\left(  \mathbf{r}\right)  \right) \label{h07}\\
&  =\frac{\sigma\sin\theta}{2}\frac{\pi}{3}+\sum_{n=1}^{+\infty}\left(
\frac{\sinh(2\pi\gamma_{n0})}{n\left[  \cosh(2\pi\gamma_{n0})-\cos(2\pi
\beta_{n})\right]  }-\frac{1}{n}\right) \nonumber\\
&  -\lim_{\mathbf{r}\rightarrow0}\left(  \frac{1}{2}\ln\left[  \left(  2\pi
t_{2}\sigma\sin\theta\right)  ^{2}+\left(  2\pi t\right)  ^{2}\right]
-\frac{1}{2}\ln\left[  r_{1}^{2}+r_{2}^{2}+2r_{1}r_{2}\cos\theta\right]
\right) \nonumber\\
&  =\frac{\sigma\sin\theta}{2}\frac{\pi}{3}-\ln\left(  2\pi l_{1}\right)
-\sum_{n=1}^{+\infty}\left(  \frac{\exp(-2\pi\gamma_{n0})-\cos(2\pi\beta_{n}%
)}{n\left[  \cosh(2\pi\gamma_{n0})-\cos(2\pi\beta_{n})\right]  }\right)
.\nonumber
\end{align}

\section{Coulomb interaction in 3D}

The Poisson equation to be solved in this case is
\begin{equation}
\nabla^{2}G(\mathbf{r})=-4\pi\sum_{l}\delta(\mathbf{r}+\mathbf{l})+\frac{4\pi
}{V},\label{h08}%
\end{equation}
where%
\begin{equation}
V=l_{1}l_{2}l_{3}\left[  \mathbf{e}_{i}\mathbf{.}(\mathbf{e}_{j}%
\times\mathbf{e}_{k})\right]  \label{hh0}%
\end{equation}
stands for the volume of the unit cell. The last term in Eq.(\ref{h08})
amounts to the presence of uniform background charge. The solution of
Eq.(\ref{h08}) is given by
\begin{equation}
G(\mathbf{r})=\frac{4\pi}{V}\lim_{\xi\rightarrow0}\left(  \sum_{\mathbf{Q}%
}\frac{\exp(i\mathbf{Q}\cdot\mathbf{r})}{\mathbf{Q}^{2}+\xi^{2}}-\frac{1}%
{\xi^{2}}\right)  ,\label{expansion1}%
\end{equation}
where%
\begin{equation}
\mathbf{r}=r_{1}\mathbf{e}_{1}+r_{2}\mathbf{e}_{2}+r_{3}\mathbf{e}%
_{3},,\,\,\,\,\,\mathbf{Q}=n_{1}\mathbf{b}_{1}+n_{2}\mathbf{b}_{2}%
+n_{3}\mathbf{b}_{3},\label{h09}%
\end{equation}
where $\mathbf{Q}$ runs over all reciprocal lattice vectors spanned by
\begin{equation}
\mathbf{b}_{i}=\frac{2\pi}{l_{i}}\frac{\mathbf{e}_{j}\times\mathbf{e}_{k}%
}{\mathbf{e}_{i}\mathbf{.}(\mathbf{e}_{j}\times\mathbf{e}_{k})},\label{h10}%
\end{equation}
for all cyclic permutations of $(i,j,k)$ and $n_{1}$,$n_{2}$ and $n_{3}$ range
from $-\infty$ to $+\infty$. Using Eqs.(\ref{expansion1}),(\ref{h09}) and
(\ref{h10}) we obtain
\begin{align}
G(\mathbf{r}) &  =\frac{4\pi}{Vb_{3}^{2}}\sum_{n_{1},n_{2},n_{3}}\frac
{\exp\left[  i2\pi\left(  n_{1}\frac{r_{1}}{l_{1}}+n_{2}\frac{r_{2}}{l_{2}%
}+n_{3}\frac{r_{3}}{l_{3}}\right)  \right]  }{\left(  n_{3}^{2}+n_{2}%
^{2}c_{22}+n_{1}^{2}c_{11}+2n_{1}n_{2}c_{12}+2n_{2}n_{3}c_{23}+2n_{3}%
n_{1}c_{31}+\xi^{2}\right)  }\label{gsum3}\\
&  -\frac{4\pi}{Vb_{3}^{2}}\frac{1}{\xi^{2}},\nonumber
\end{align}
where $0\leq r_{i}/l_{i}<1$,%

\begin{equation}
c_{ij}=\frac{\mathbf{b}_{i}.\mathbf{b}_{j}}{\mathbf{b}_{3}.\mathbf{b}_{3}%
}\text{ \ \ \ \ \ \ \ \ \ \ \ \ \ \ \ \ }1\leq i,j\leq3, \label{h11}%
\end{equation}
and we have, as before, introduced an infinitesimal parameter $\xi$ in the
denominator and subtracted a counter term from the whole sum due to the
presence of a uniform background charge. We now evaluate the sum
\begin{equation}
L(n_{1},n_{2,}r_{3},\xi)=\frac{1}{\pi}\sum_{n_{3}=-\infty}^{\infty}\frac{\exp\left(
i\frac{2\pi}{l_{3}}n_{3}r_{3}\right)  }{\left(  n_{3}^{2}+n_{2}^{2}%
c_{23}+n_{1}^{2}c_{13}+2n_{1}n_{2}c_{12}+2n_{2}n_{3}c_{23}+2n_{3}n_{1}%
c_{31}+\xi^{2}\right)  }. \label{h12}%
\end{equation}
This sum can be obtained easily and the result is
\begin{align}
&  L(n_{1},n_{2,}r_{3},\xi)\label{l}\\
&  =\exp(i2\pi\beta_{n_{1},n_{2}}t_{3})\frac{\exp\left[  -i2\pi\beta
_{n_{1},n_{2}}\right]  \sinh(2\pi\gamma_{n_{1},n_{2}}t_{3})+\sinh\left[
2\pi\gamma_{n_{1},n_{2}}(1-t_{3})\right]  }{\gamma_{n_{1},n_{2}}\left[
\cosh(2\pi\gamma_{n_{1},n_{2}})-\cos(2\pi\beta_{n_{1},n_{2}})\right]
}\nonumber
\end{align}
where%

\begin{equation}
t_{i}=\frac{r_{i}}{l_{i}}\text{ \ \ \ \ \ \ \ \ for }i=1,2\text{ and
}3,\label{a8}%
\end{equation}
\begin{equation}
\beta_{n_{1},n_{2}}=-n_{1}c_{31}-n_{2}c_{32},\label{h13}%
\end{equation}
and%

\begin{equation}
\gamma_{n_{1},n_{2}}=\left[  n_{2}^{2}c_{22}+n_{1}^{2}c_{11}+2 n_{1} n_{2}
c_{12}-\left(  n_{1}c_{31}+n_{2}c_{32}\right)  ^{2}+\xi^{2}\right]  ^{1/2}.
\label{h14}%
\end{equation}
Plugging the value of $L(n_{1},n_{2,}r_{3},\xi)$ from Eq. (\ref{l}) in Eq.
(\ref{gsum3}) we obtain,%

\begin{align}
G(\mathbf{r}) &  =\frac{4\pi^2}{Vb_{3}^{2}}\sum_{n_{1},n_{2}}\exp\left[
i2\pi(n_{1}t_{1}+n_{2}t_{2})\right]  ~L(n_{1},n_{2,}r_{3},\xi)-\frac{4\pi
}{Vb_{3}^{2}}\frac{1}{\xi^{2}}\label{h15}\\
&  =\frac{4\pi^2}{Vb_{3}^{2}}\left.  \sum_{n_{1},n_{2}}^{\prime}\exp\left[
i2\pi(n_{1}t_{1}+n_{2}t_{2})\right]  ~L\left(  n_{1},n_{2,}r_{3},\xi\right)
\right\vert _{\xi=0}\nonumber\\
&  +\frac{4\pi^2}{Vb_{3}^{2}}\frac{\pi}{3}\left(  1-6t_{3}+6t_{3}^{2}\right)
~,\nonumber
\end{align}
where a prime over summation sign implies $n_{1}$ and $n_{2}$ cannot both be
zero simultaneously. We have also separated out the term corresponding to
$n_{1}=0$ and $n_{2}=0$ in Eq.(\ref{h15}) and taken the limit $\xi
\rightarrow0$, which results in cancellation of the diverging factor
$4\pi/\left(  Vb_{3}^{2}\xi^{2}\right)  $. Eq. (\ref{h15}) is one of the main
results of this paper. It is easy to see that the sum defined in
Eq.(\ref{h15}) fails to converge fast enough as $t_{i}$ tend to zero. In fact,
towards large values of $\gamma_{n_{1},n_{2}}$, the quantity $L$ defined in
Eq.(\ref{h15}) goes as $\exp\left(  -2\pi\gamma_{n_{1},n_{2}}t_{3}\right)  $
and if $t_{3}$ is small, this convergence may be very slow. As before for the
2D case, we only concentrate on that part of unit cell which corresponds to
$0\leq t_{i}\leq0.5$. To transform Eq. (\ref{h15}) in a form, which converges
even for small values of $t_{i}$, we need to separate out a term which
corresponds to slab geometry. By slab geometry we mean a situation which is
obtained by sending one of the sides of the unit cell to infinity. We write
\begin{align}
G(\mathbf{r}) &  =\frac{4\pi^2}{Vb_{3}^{2}}\left.  \sum_{n_{1},n_{2}}^{\prime
}\exp\left[  i2\pi(n_{1}t_{1}+n_{2}t_{2})\right]  ~B\left(  n_{1},n_{2,}%
r_{3},\xi\right)  \right\vert _{\xi=0}+G_{2}(\mathbf{r})\label{h16}\\
&  +\frac{4\pi^2}{Vb_{3}^{2}}\frac{\pi}{3}\left(  1+6t_{3}^{2}\right)  ,
\end{align}
where $G_{2}$ corresponds to the slab geometry case and is given by, %

\begin{align}
G_{2}(\mathbf{r})  &  =\frac{4\pi^2}{Vb_{3}^{2}}\sum_{n_{1},n_{2}}^{\prime}%
\frac{\exp\left[  i2\pi\beta_{n_{1},n_{2}}t_{3}+i2\pi(n_{1}t_{1}+n_{2}%
t_{2})\right]  \exp(-2\pi\gamma_{n_{1},n_{2}}t_{3})}{\gamma_{n_{1},n_{2}}%
}\label{h17}\\
&  -\frac{4\pi^2}{Vb_{3}^{2}}\frac{\pi}{3}\left(  6t_{3}\right)  ,\nonumber
\end{align}
and $B$ is defined as%

\begin{equation}
B(n_{1},n_{2,}r_{3},\xi)=L(n_{1},n_{2,}r_{3},\xi)-\frac{\exp(i2\pi
\beta_{n_{1},n_{2}}t_{3})\exp(-2\pi\gamma_{n_{1},n_{2}}t_{3})}{\gamma
_{n_{1},n_{2}}}.\label{h18}%
\end{equation}
The result in Eq.(\ref{h17}) represents Coulomb interaction with open boundary
condition along the $r_{3}$ direction and periodic boundaries along the
$r_{1}$ and $r_{2}$ directions. $G_{2}$ can be obtained from $G$ by taking the
limit $l_{3}\rightarrow\infty$ and dropping a constant term. We also note
above that both $\left(  \beta_{n_{1},n_{2}}t_{3}\right)  $ and $\left(
\gamma_{n_{1},n_{2}}t_{3}\right)  $ are independent of $l_{3},$ when
$\xi\rightarrow0$. 

The term in Eq.(\ref{h18}), can be written as
\begin{align}
&  B(n_{1},n_{2,}r_{3},\xi)\label{h19}\\
&  =-\frac{\exp\left(  i2\pi\beta_{n_{1},n_{2}}t_{3}\right)  \left[
\cosh\left[  2\pi\left(  i\beta_{n_{1},n_{2}}-\gamma_{n_{1},n_{2}}%
t_{3}\right)  \right]  -\exp(-2\pi\gamma_{n_{1},n_{2}})\cosh(2\pi
\gamma_{n_{1},n_{2}}t_{3})\right]  }{\gamma_{n_{1},n_{2}}\left[  \cosh
(2\pi\gamma_{n_{1},n_{2}})-\cos(2\pi\beta_{n_{1},n_{2}})\right]  }.\nonumber
\end{align}
It can be easily seen that, for large $\gamma_{n_{1},n_{2}}$, the slowest
decaying term on the right hand side of Eq.(\ref{h19}) goes as $\exp\left[
-2\pi\gamma_{n_{1},n_{2}}(1-t_{3})\right]  $. So, the fastest convergence now
occurs for $t_{3}=0$ and slowest for $t_{3}=0.5$. But even this `slowest'
convergence for $t_{3}=0.5$, \ amounts to an extremely fast exponential
convergence of $\exp(-\pi\gamma_{n_{1},n_{2}})$.

Essentially now the whole problem has reduced to a fast evaluation of $G_{2}$
in Eq.(\ref{h17}). We take up this case now. As $G_{2}(\mathbf{r})$ fails to
converge fast enough for small separations, we break the sum in Eq.(\ref{h19})
into two parts%

\begin{equation}
\sum_{n_{1},n_{2}}^{\prime}=\sum_{n_{1},n_{2}^{\prime}}+\sum_{n_{1},n_{2}%
=0}^{\prime}.
\end{equation}
Note that in our notation $\sum_{n}^{\prime}$ and $\sum_{n^{\prime}}$
represent the same thing. Thus $G_{2}(\mathbf{r})$ can be written as%

\begin{equation}
G_{2}(\mathbf{r})=G_{2}^{\prime}(\mathbf{r})+G_{20}(\mathbf{r}), \label{h20}%
\end{equation}
where
\begin{equation}
G_{2}^{\prime}(\mathbf{r})=\frac{4\pi^2}{Vb_{3}^{2}}\sum_{n_{2}}^{\prime}%
\exp\left(  i2\pi n_{2}x_{2}\right)  \left(  \sum_{n_{1}=-\infty}^{\infty
}\frac{\exp\left(  2\pi in_{1}x_{1}\right)  \exp(-2\pi\gamma_{n_{1},n_{2}%
}t_{3})}{\gamma_{n_{1},n_{2}}}\right)  . \label{h21}%
\end{equation}
and%

\begin{align}
G_{20}(\mathbf{r}) &  =\frac{4\pi^2}{Vb_{3}^{2}}\sum_{n_{1}}^{\prime}\left.
\frac{\exp\left[  i2\pi n_{1}\left(  -c_{31}t_{3}+t_{1}\right)  \right]
\exp(-2\pi\gamma_{n_{1},n_{2}}t_{3})}{\gamma_{n_{1},n_{2}}}\right\vert
_{n_{2}=0,~\xi=0}\label{h22}\\
&  -\frac{4\pi^2}{Vb_{3}^{2}}2\pi t_{3}.\nonumber
\end{align}
To further transform Eq. (\ref{h22}), we express $\gamma_{n_{1},n_{2}}$ as
\begin{align}
\gamma_{n_{1},n_{2}} &  =\left[  n_{1}^{2}\left(  c_{11}-c_{13}^{2}\right)
+n_{2}^{2}\left(  c_{22}-c_{23}^{2}\right)  +2n_{1}n_{2}(c_{12}-c_{13}%
c_{23})+\xi^{2}\right]  ^{1/2}\label{h25}\\
&  =\left(  \left[  n_{1}\widetilde{\delta}+n_{2}\widetilde{a}\right]
^{2}+n_{2}^{2} \widetilde{b}^{2}\right)  ^{1/2},\nonumber
\end{align}
where we have put $\xi=0$ and
\begin{align}
\widetilde{\delta} &  =\left(  c_{11}-c_{13}^{2}\right)  ^{1/2}%
,\,\,\,\widetilde{a}=\frac{\left(  c_{12}-c_{13}c_{23}\right)  }%
{\widetilde{\delta}},\,\,\label{h26}\\
\widetilde{b} &  =\frac{\left[  \left(  c_{11}-c_{13}^{2}\right)  \left(
c_{22}-c_{23}^{2}\right)  -\left(  c_{12}-c_{13}c_{23}\right)  ^{2}\right]
^{1/2}}{\widetilde{\delta}}.\nonumber
\end{align}
As the convergence of Eq.(\ref{h16}) crucially depends on the value of
$\gamma_{n_{1},n_{2}},$ it is helpful at this point to note that the minimum
value of $\gamma_{n_{1},n_{2}}$ , when both $n_{1}$ and $n_{2}$ are integers
such that they cannot both be zero simultaneously, is given by%
\begin{equation}
\gamma_{\text{min}}^{2}=\widetilde{b}^{2}\min\left(  1,\frac{\widetilde
{\delta}^{2}}{\widetilde{a}^{2}+\widetilde{b}^{2}}\right)  .
\end{equation}
We note that $\gamma_{\text{min}}$ depends upon the geometry of the unit cell.
To get a fast convergence, it is imperative that the sides of the triclinic
unit cell are labelled such that $\gamma_{\text{min}}$ is as large as possible.

Now, we consider the sum $G_{20}$ defined in Eq.(\ref{h22}). Using the
relation,%
\begin{equation}
Vb_{3}^{2}\left\vert \widetilde{\delta}\right\vert =4\pi^2 l_{2},
\end{equation}
which is derived in Appendix B, we can write%
\begin{align}
G_{20}(\mathbf{r}) &  =\frac{\left\vert \widetilde{\delta}\right\vert }{l_{2}%
}\sum_{n_{1}}^{\prime}\frac{\exp\left[  -2\pi n_{1}t_{3} \left\vert
\widetilde{\delta}\right\vert \right]  }{\left\vert n_{1}\right\vert
\left\vert \widetilde{\delta}\right\vert }\exp\left(  2\pi i n_{1}%
x_{1}\right)  -\frac{2\pi~t_{3}\left\vert \widetilde{\delta}\right\vert
}{l_{2}}\\
&  =-\frac{1}{l_{2}}\ln\left[  \cosh\left(  2\pi t_{3}\widetilde{\delta
}\right)  -\cos\left( 2\pi x_{1}\right)  \right]  -\frac{\ln\left(
2\right)  }{l_{2}},\nonumber
\end{align}
where%

\begin{equation}
x_{1}=-c_{31}t_{3}+t_{1},~x_{2}=-c_{32}t_{3}+t_{2},
\end{equation}
and we have used the identity from Eq.(\ref{h03}). Thus, we have been able to
obtain $G_{20}(\mathbf{r})$ analytically. Now, we transform $G_{2}^{\prime
}(\mathbf{r})$. The sum over $n_{1}$ in Eq.(\ref{h21}) ,%

\begin{equation}
S\left(  n_{2},r_{1},r_{3}\right)  =\sum_{n_{1}=-\infty}^{\infty}\frac
{\exp\left(  2\pi in_{1}x_{1}\right)  \exp(-2\pi\gamma_{n_{1},n_{2}}t_{3}%
)}{\gamma_{n_{1},n_{2}}}, \label{h27}%
\end{equation}
can be transformed using an identity,%

\begin{align}
&  \sum_{n}\frac{\exp\left(  -\beta\sqrt{\alpha^{2}+\left(  q+n \delta
\right)  ^{2}}\right)  }{\sqrt{\alpha^{2}+\left(  q+n \delta\right)  ^{2}}%
}\exp\left[  i p\left(  q+n \delta\right)  \right] \label{h28}\\
&  =\frac{2}{\left\vert \delta\right\vert }\sum_{n}K_{0}\left(  \alpha
\sqrt{\beta^{2}+\left(  2\pi\frac{n}{\delta}-p\right)  ^{2}}\right)
\exp\left(  2\pi i\frac{n}{\delta}q\right)  ,\nonumber
\end{align}
which can be derived with a simple application of Jacobi Poisson theorem
\cite{hautot} to the integral%

\begin{equation}
K_{0}\left(  a\sqrt{b^{2}+x^{2}}\right)  =\frac{1}{2}\int_{-\infty}^{+\infty
}dy\frac{\exp\left(  -b\sqrt{a^{2}+y^{2}}\right)  }{\sqrt{a^{2}+y^{2}}}%
\exp\left(  ixy\right)  . \label{h29}%
\end{equation}
Identifying%

\begin{equation}
\beta=2\pi t_{3},~p=2\pi \frac{x_{1}}{\widetilde{\delta}},~q=n_{2}\widetilde{a}%
,~\alpha=\left\vert n_{2}\widetilde{b}\right\vert \text{ and }\delta
=\widetilde{\delta}, \label{h30}%
\end{equation}
one obtains
\begin{align}
S\left(  n_{2},r_{1},r_{3}\right)  =  &  \exp\left(  -2\pi i \frac{x_{1}%
}{\widetilde{\delta}}n_{2}\widetilde{a}\right)  \sum_{n_{1}=-\infty}^{\infty
}\frac{\exp\left(  -2\pi t_{3}\sqrt{\left\vert n_{2}\widetilde{b}\right\vert
^{2}+\left(  n_{2}\widetilde{a}+n_{1}\widetilde{\delta}\right)  ^{2}}\right)
}{\sqrt{\left\vert n_{2}\widetilde{b}\right\vert ^{2}+\left(  n_{2}%
\widetilde{a}+n_{1}\widetilde{\delta}\right)  ^{2}}}\label{h31}\\
&  \times\exp\left[2\pi i \frac{x_{1}}{\widetilde{\delta}}\left(  n_{2}%
\widetilde{a}+n_{1}\widetilde{\delta}\right)  \right] \nonumber\\
=\frac{2}{\left\vert \widetilde{\delta}\right\vert }  &  \exp\left(
-2\pi i \frac{x_{1}}{\widetilde{\delta}}n_{2}\widetilde{a}\right)  \times
\sum_{n_{1}}K_{0}\left(  2\pi\left\vert n_{2}\widetilde{b}\right\vert
\sqrt{t_{3}^{2}+\left(  \frac{n_{1}-x_{1}}{\widetilde{\delta}}\right)  ^{2}%
}\right) \nonumber\\
&  \times\exp\left(  2\pi i\frac{n_{1}}{\widetilde{\delta}}n_{2}\widetilde
{a}\right)  .\nonumber
\end{align}
Substituting the value of $S\left(  n_{2},r_{1},r_{3}\right)  $ in
Eq.(\ref{h21}) we obtain%

\begin{align}
G_{2}^{\prime}(\mathbf{r}) &  =\frac{2}{l_{2}}\sum_{n_{2}}^{\prime}\exp\left[
2\pi in_{2}\left(  x_{2}-\frac{x_{1}\widetilde{a}}{\widetilde{\delta}}\right)
\right]  \label{h32}\\
&  \times\sum_{n_{1}}K_{0}\left(  2\pi\left\vert n_{2}\widetilde{b}\right\vert
\sqrt{t_{3}^{2}+\left(  \frac{n_{1}-x_{1}}{\widetilde{\delta}}\right)  ^{2}%
}\right)  \exp\left(  2\pi i\frac{n_{1}}{\widetilde{\delta}}n_{2}\widetilde
{a}\right)  .\nonumber
\end{align}
Combining Eqs.(\ref{h20}), (\ref{h22}) and (\ref{h32}) we get one of the main
results of this paper,%

\begin{align}
G_{2}(\mathbf{r}) &  =\frac{2}{l_{2}}\sum_{n_{2}}^{\prime}\exp\left[
  2\pi i n_{2}\left(  x_{2}-\frac{x_{1}\widetilde{a}}{\widetilde{\delta}}\right)
\right]  \label{e09}\\
&  \times\sum_{n_{1}}K_{0}\left(  2\pi\left\vert n_{2}\widetilde{b}\right\vert
\sqrt{t_{3}^{2}+\left(  \frac{n_{1}-x_{1}}{\widetilde{\delta}}\right)  ^{2}%
}\right)  \exp\left(  2\pi i\frac{n_{1}}{\widetilde{\delta}}n_{2}\widetilde
{a}\right)  \nonumber\\
&  -\frac{1}{l_{2}}\ln\left[  \cosh\left(  2\pi t_{3}\widetilde{\delta
}\right)  -\cos\left( 2\pi x_{1}\right)  \right]  -\frac{\ln\left(
2\right)  }{l_{2}}.\nonumber
\end{align}
Result in Eq.(\ref{e09}) represents the sum for slab geometry and generalizes
the results of Arnold \textit{et al}\cite{arnold}. Similar expressions were
obtained by Liem \textit{et al.}\cite{liem}. Substitution of $G_{2}$ from
Eq.(\ref{e09}) in Eq.(\ref{h16}) gives us an alternative form of $G$. We note
that the problem of convergence with Eq.(\ref{e09}) still persists if the
charges are close together. The slowest converging term in Eq.(\ref{e09}) goes
as $K_{0}\left(  2\pi\left\vert n_{2}\right\vert \rho\right) $ and it
still does not converge fast enough when,
\begin{equation}
\rho=\widetilde{b}\sqrt{t_{3}^{2}+\left(  \frac{x_{1}}{\widetilde{\delta}%
}\right)  ^{2},}\label{h33}%
\end{equation}
is small. The problem of convergence lies with only those terms corresponding
to $n_{1}=0$. So, we separate out these terms%

\begin{equation}
G_{2}^{\prime}=G_{2}^{\prime\prime}+G_{1d}, \label{h34}%
\end{equation}
where
\begin{align}
G_{2}^{\prime\prime}  &  =\frac{2}{l_{2}}\sum_{n_{2}}^{\prime}\exp\left[  2\pi
in_{2}\left(  x_{2}-\frac{x_{1}\widetilde{a}}{\widetilde{\delta}}\right)
\right] \label{h35}\\
&  \times\sum_{n_{1}}^{\prime}K_{0}\left(  2\pi\left\vert n_{2}\widetilde
{b}\right\vert \sqrt{t_{3}^{2}+\left(  \frac{n_{1}-x_{1}}{\widetilde{\delta}%
}\right)  ^{2}}\right)  \exp\left(  2\pi i\frac{n_{1}}{\widetilde{\delta}%
}n_{2}\widetilde{a}\right) \nonumber
\end{align}
and%

\begin{align}
G_{1d}  &  =\frac{2}{l_{2}}\sum_{n_{2}}^{\prime}\exp\left[  2\pi in_{2}\left(
x_{2}-\frac{x_{1}\widetilde{a}}{\widetilde{\delta}}\right)  \right]
~K_{0}\left(  2\pi\left\vert n_{2}\widetilde{b}\right\vert \sqrt{t_{3}%
^{2}+\left(  \frac{x_{1}}{\widetilde{\delta}}\right)  ^{2}}\right)
\label{h36}\\
&  =\frac{4}{l_{2}}\sum_{n_{2}=1}^{\infty}\cos\left[  2\pi n_{2}\left(
x_{2}-\frac{x_{1}\widetilde{a}}{\widetilde{\delta}}\right)  \right]
\,K_{0}\left(  2\pi n_{2}\widetilde{b}\sqrt{t_{3}^{2}+\left(  \frac{x_{1}%
}{\widetilde{\delta}}\right)  ^{2}}\right)  .\nonumber
\end{align}
The term $G_{2}^{\prime\prime}$ does not have any convergence problem for
small separation between the two charges. We need to apply a final
transformation to the sum in Eq.(\ref{h36}). We start with the identity
\cite{sperb2},%
\begin{align}
f\left(  \rho,x \right)   &  =4\sum_{m=1}^{\infty} K_{0}\left(  2\pi m
\rho\right)  \cos\left(  2\pi m x \right) \label{h37}\\
&  =2\left\{  \gamma+\ln\left(  \frac{\rho}{2}\right)  \right\}  +\frac
{1}{\sqrt{\rho^{2}+x^{2}}}\nonumber\\
&  +\sum_{n_{1}=1}^{N-1}\left(  \frac{1}{\sqrt{\rho^{2}+\left(  n_{1}%
+x\right)  ^{2}}}+\frac{1}{\sqrt{\rho^{2}+\left(  n_{1}-x\right)  ^{2}}%
}\right) \nonumber\\
&  -2\gamma-\left\{  \psi(N+x)+\psi(N-x)\right\} \nonumber\\
&  +\sum_{l=1}^{\infty}\binom{-1/2}{l}\rho^{2l}\left(  \zeta\left(
2l+1,N+x\right)  +\zeta\left(  2l+1,N-x\right)  \right)  ,\nonumber
\end{align}
where $\psi$ and $\zeta$ stand for digamma and Hurwitz Zeta function
respectively. $N\geq1$ is the smallest integer chosen such that it satisfies
the condition $N>\rho+x$. However for better convergence it is desirable that
one chooses $N$ such that $N>\rho+1$. Now, identifying $\rho$ from
Eq.(\ref{h33}) and
\begin{equation}
x=\left\vert \left(  x_{2}-\frac{x_{1}\widetilde{a}}{\widetilde{\delta}%
}\right)  \right\vert \label{h38}%
\end{equation}
and realizing that (see Appendix B)
\begin{equation}
\rho^{2}+x^{2}=\frac{r_{1}^{2}+r_{2}^{2}+r_{3}^{2}+2r_{1}r_{2}\cos
\alpha+2r_{2}r_{3}\cos\beta+2r_{3}r_{1}\cos\gamma}{l_{2}^{2}}, \label{h39}%
\end{equation}
one obtains an expression for $G$, which converges exponentially fast even for
small $x_{i}$:%

\begin{align}
G(\mathbf{r}) &  =\frac{\left\vert \widetilde{\delta}\right\vert }{l_{2}%
}\left.  \sum_{n_{1},n_{2}}^{\prime}\exp\left[  2\pi i(n_{1}t_{1}+n_{2}%
t_{2})\right]  ~B(n_{1},n_{2,}r_{3},\xi)\right\vert _{\xi=0}\label{h40}\\
&  +\frac{2}{l_{2}}\left\{  \sum_{n_{1}^{\prime},n_{2}^{\prime}}\exp\left[
2\pi in_{2}\left(  x_{2}-\frac{x_{1}\widetilde{a}}{\widetilde{\delta}}\right)
\right]  \right.  \nonumber\\
&  \times\left.  K_{0}\left(  2\pi\left\vert n_{2}\text{ }\widetilde
{b}\right\vert \sqrt{t_{3}^{2}+\left(  \frac{n_{1}-x_{1}}{\widetilde{\delta}%
}\right)  ^{2}}\right)  \exp\left(  2\pi i\frac{n_{1}}{\widetilde{\delta}%
}n_{2}\widetilde{a}\right)  \right\}  \nonumber\\
&  -\frac{1}{l_{2}}\ln\left[  \cosh\left(  2\pi t_{3}\widetilde{\delta
}\right)  -\cos\left(  2\pi x_{1}\right)  \right]  -\frac{\ln\left(  2\right)
}{l_{2}}+\frac{2\gamma}{l_{2}}\nonumber\\
&  +\frac{\left\vert \widetilde{\delta}\right\vert }{l_{2}}\frac{\pi}%
{3}\left(  1+6t_{3}^{2}\right)  +\frac{2}{l_{2}}\ln\left(  \frac{\rho}%
{2}\right)  -\frac{\psi(N+x)+\psi(N-x)}{l_{2}}\nonumber\\
&  +\frac{1}{l_{2}}\sum_{n=1}^{\infty}\binom{-1/2}{n}\rho^{2n}\left[
\zeta\left(  2n+1,N+x\right)  +\zeta\left(  2n+1,N-x\right)  \right]
\nonumber\\
&  +\frac{1}{l_{2}}\sum_{n_{1}=1}^{N-1}\left(  \frac{1}{\sqrt{\rho^{2}+\left(
n_{1}+x\right)  ^{2}}}+\frac{1}{\sqrt{\rho^{2}+\left(  n_{1}-x\right)  ^{2}}%
}\right)  \nonumber\\
&  +\frac{1}{\left(  r_{1}^{2}+r_{2}^{2}+r_{3}^{2}+2r_{1}r_{2}\cos
\alpha+2r_{2}r_{3}\cos\beta+2r_{3}r_{1}\cos\gamma\right)  ^{1/2}}.\nonumber
\end{align}
Even though Eq.(\ref{h40}) gives a very good convergence for smaller values of
$r_{i}<\varepsilon=10^{-3}$, it is not defined when $t_{3}=0$ and $x_{1}=0$.
The problem lies in the logarithmic terms, which can be combined together such
that the opposing logarithmic divergences cancel each other as shown below.
For small separations, it can be easily shown \cite{paper3} that%
\begin{align}
\ln\left[  \cosh y-\cos x\right]   &  =\ln\left[  \frac{y^{2}+x^{2}}%
{2}\right]  +\ln\left\{  1+\frac{2!}{4!}\left(  y^{2}-x^{2}\right)  +\frac
{2!}{6!}\left(  y^{4}-x^{2}y^{2}+x^{4}\right)  \right.  \\
&  \left.  +\frac{2!}{8!}\left(  y^{4}+x^{4}\right)  \left(  y^{2}%
-x^{2}\right)  +\text{O}\left[  x^{8},y^{8}\right]  \right\}  .\nonumber
\end{align}
Thus all of the logarithmic terms in Eq.(\ref{h40}) can be combined together
as%
\begin{align}
&  -\frac{1}{l_{2}}\ln\left[  \cosh\left(  2\pi t_{3}\widetilde{\delta
}\right)  -\cos\left(  2\pi x_{1}\right)  \right]  -\frac{\ln\left(  2\right)
}{l_{2}}+\frac{2}{l_{2}}\ln\left(  \frac{\rho}{2}\right)  \label{reg}\\
&  =-\frac{2}{l_{2}}\ln\left(  \frac{4\pi\widetilde{\delta}}{\widetilde{b}%
}\right)  +\ln\left\{  1+\frac{2!}{4!}\left[  \left(  t_{3}\widetilde{\delta
}\right)  ^{2}-x_{1}^{2}\right]  +\frac{2!}{6!}\left[  \left(  t_{3}%
\widetilde{\delta}\right)  ^{4}-x_{1}^{2}\left(  t_{3}\widetilde{\delta
}\right)  ^{2}+x_{1}^{4}\right]  \right.  \nonumber\\
&  +\left.  +\frac{2!}{8!}\left[  \left(  t_{3}\widetilde{\delta}\right)
^{4}+x_{1}^{4}\right]  \left[  \left(  t_{3}\widetilde{\delta}\right)
^{2}-x_{1}^{2}\right]  +\text{O}\left[  x_{1}^{8},\left(  t_{3}\widetilde
{\delta}\right)  ^{8}\right]  \right\}  .\nonumber
\end{align}
\qquad The RHS of Eq.(\ref{reg}) remains regular even when $x_{1}$ and $t_{3}$
both tend to zero. The self-energy of the system can be easily obtained now as,%

\begin{align}
G_{\text{self}}^{\text{3D}}(\mathbf{r}) &  =\lim_{\left(  r_{1},r_{2}%
,r_{3}\right)  \rightarrow\left(  0,0,0\right)  }\left(  G(r_{1},r_{2}%
,r_{3})-\frac{1}{\left(  r_{1}^{2}+r_{2}^{2}+r_{3}^{2}+2r_{1}r_{2}\cos
\alpha+2r_{2}r_{3}\cos\beta+2r_{3}r_{1}\cos\gamma\right)  ^{1/2}}\right)
\label{h41}\\
&  =\frac{\left\vert \widetilde{\delta}\right\vert }{l_{2}}\left.  \sum
_{n_{1},n_{2}}^{\prime}~B(n_{1},n_{2,}0,\xi)\right\vert _{\xi=0}+\frac
{2}{l_{2}}\sum_{n_{1}^{\prime},n_{2}^{\prime}}K_{0}\left(  2\pi\left\vert
n_{1}n_{2}\frac{\widetilde{b}}{\widetilde{\delta}}\right\vert \right)
\times\exp\left(  2\pi i n_{1}n_{2}\frac{\widetilde{a}}{\widetilde{\delta}%
}\right)  \nonumber\\
&  +\frac{\left\vert \widetilde{\delta}\right\vert }{l_{2}}\frac{\pi}{3}%
-\frac{2}{l_{2}}\ln\left(  \frac{4\pi\widetilde{\delta}}{\widetilde{b}%
}\right)  +\frac{2\gamma}{l_{2}}.\nonumber
\end{align}
We have thus obtained complete expressions for $G$ and the self-energy.

\section{Results and Conclusions}

We have obtained complete expressions for the logarithmic potential in 2D and
Coulomb potential in 3D, including the self-energies. The results were derived
for most general cases, that is a rhombic cell in 2D and a triclinic cell in
3D. To my knowledge, this is the first time a practical method has been
developed in 3D, which is different from the Ewald method, and yet may be
applied to a triclinic unit cell to obtain periodic Coulomb sums. Even though
the formulas developed here look complicated, their implementation on a
computer will be marginally difficult from the case of orthorhombic unit cell.
The formulas derived here converge extremely fast and require only a few dozen
terms at worst to obtain results to a very high accuracy as opposed to the
Ewald method, which may require close to 200 to 300 terms for the same
calculations. In the process, we have simplified and solved a problem
mentioned by Crandall\cite{crandall}, that of finding Coulomb potential in
close vicinity of a particle under the PBC. An important implication of the
formulas derived here is that most part of the interaction can be calculated
linearly in the number of charges present in the system. For more details on
how this can be achieved, we refer the reader to Sperb who discusses this in
the context of an orthorhombic cell. The results obtained in this paper reduce
to the results of a recent paper\cite{paper3} when all angles pertaining to
the unit cell are set to $\pi/2$.

The results for 3D triclinic case may be obtained by directly generalizing
Lekner's work. This work by the author will be presented elsewhere. We also
note that the logarithmic sum in 2D for a rhombic cell may be obtained in a
closed form. This will be the subject of another paper. Also, here we would
like to point out a connection between the results of slab geometry and that
of 3D triclinic cell. As it has been shown here, the 3D Green function can be
broken in two parts. The first part corresponds to the slab geometry Green
function and the second part takes into account the rest of the layers. Thus
following Ref.\onlinecite{arnold1} one can make use of this relation to obtain
potential energies for the slab geometry cases by employing the result for the
triclinic cell.

A naive application of most methods gives a scaling which goes as $N^{2}$,
where $N$ is the number of charges in the unit cell. However, the Ewald method
can be optimized\cite{ceperley} to give a scaling of $[N\ln(N)]^{3/2}$.
Strebel \textit{et
al.\cite{strebel}} gave an approximate method, which they call the MMM method,
which is based on the formulas developed by Sperb in his earlier
work\cite{sperb2}. With the help of MMM one can achieve a $N\ln(N)$ scaling.
There is another approximate method in use to achieve a faster scaling. This
method is known as PPPM. In Ref.\onlinecite{strebel}, however, it was shown
that for $N>2^{10}$ and a relative tolerance of $10^{-4}$ the MMM is the best
method available. As the formulas derived here are a generalization of Sperb's
work, it may now be possible, by using the results presented in this work, to
employ the MMM method to achieve a scaling of $N\ln(N)$ even for a triclinic
cell. Similarly for the logarithmic interaction in 2D, it should be possible
to achieve $N\ln(N)$ scaling.

In short, we believe the method developed here is an alternative to the Ewald
method. The formulas developed here generalize and simplify
Sperb's\cite{sperb2} work. From the results of summation formula derived for a
triclinic cell, it is easy to obtain results for $2D+h$ slab geometry and vice
versa. For the slab geometry case expressions obtained here generalize the
work of Arnold \textit{et al}\cite{arnold}\textit{.} and give an alternate
derivation of the results obtained by Liem \textit{et al.}\cite{liem}. The
formulas derived in this work can be easily employed to calculate the Madelung
potential for any periodic crystal in 3D, where a triclinic cell repeats
itself to infinity under the PBC.

\acknowledgments{ I am thankful to Dr. Y. Y. Goldschmidt for useful discussions. I also
thank Barun K. Dhar and Mahesh Bandi for suggesting several
improvements in the paper.}

\appendix

\section{Complex Sum}

The usual way to sum over series of type
\begin{equation}
S=\sum_{n=-\infty}^{\infty}f\left(  n\right)
\end{equation}
is to consider the integral
\begin{equation}
I=%
{\displaystyle\oint}
f\left(  z\right)  \pi\cot\left(  \pi z\right)  dz. \label{c1}%
\end{equation}
It is required that the function $f\left(  z\right)  $ satisfies the condition
that integral $I$ becomes zero when the contour of integration is chosen to be
$C_{N}$ as shown in Fig.2. 

\begin{figure}[ptb]
\begin{center}
\includegraphics[angle=0,scale=0.6]{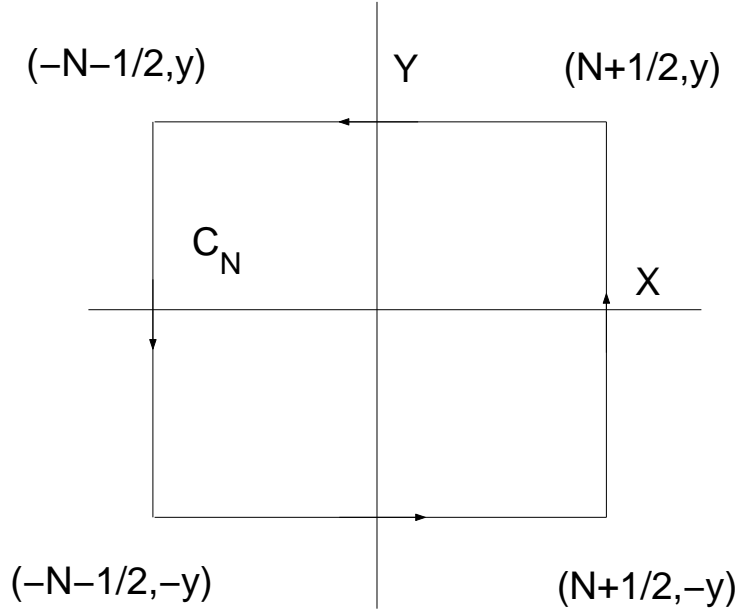}
\end{center}
\label{contour}\caption{Contour of integration, $C_{N}$. Both $N$ and $y$ tend
to infinity.}%
\end{figure}

The poles of $\pi\cot\left(  \pi z\right)  $ fall
at $z=n$ where $n=0,\pm1,\pm2,...$. Then by residue theorem we have
\begin{equation}
\sum_{n=-\infty}^{\infty}f\left(  n\right)  =-\text{sum of residues of
}f\left(  z\right)  \pi\cot\left(  \pi z\right)  \text{ at the poles of
}f\left(  z\right)  .
\end{equation}
Here, in particular, we consider the function%
\begin{equation}
f\left(  n\right)  =\frac{\exp\left(  in\alpha\right)  }{\left(
n-\beta\right)  ^{2}+\gamma^{2}}\text{ }\ \ \ \ \ \ \ \ \ \alpha<2\pi,
\label{c2}%
\end{equation}
where $x$, $\beta$, $\gamma$ are real numbers and $x>0$. The results obtained
here are more general in nature and may be applied for other forms of $f(n)$.
It can be easily verified that the function given above does not satisfy the
condition that integral $I$ go to zero for contour $C_{N}$. A trick which is
usually not found in books may help solve the problem. Instead of considering
the integral $I$ in Eq.(\ref{c1}), we consider the following integral%
\begin{equation}
I^{\prime}=%
{\displaystyle\oint}
f\left(  z\right)  \left(  -1\right)  ^{z}\pi\csc\left(  \pi z\right)  dz,
\end{equation}
where have in mind that $\exp\left(  -i\pi\right)  =-1$. Residues of $f\left(
z\right)  (-1)^{z}\pi\csc\left(  \pi z\right)  $ at $z=n$, $n=0,\pm
1,\pm2,...,$ is%
\begin{equation}
\lim_{z\rightarrow n}f\left(  z\right)  \left(  z-n\right)  (-1)^{z}\pi
\csc\left(  \pi z\right)  =f\left(  n\right)  .
\end{equation}
Thus if the integral $I^{\prime}$ goes to zero for the contour $C_{N}$ then we
obtain%
\begin{equation}
\sum_{n=-\infty}^{\infty}f\left(  n\right)  =-\text{sum of residues of
}f\left(  z\right)  (-1)^{z}\pi\csc\left(  \pi z\right)  \text{ at the poles
of }f\left(  z\right)  . \label{a2}%
\end{equation}
Function $f$ given in Eq.(\ref{c2}) does satisfy the condition that
$I^{\prime}=0$ when the integration is evaluated for the contour $C_{N}$. To
show this, we concentrate on the the following function%
\begin{equation}
g\left(  x,y,\alpha\right)  =\frac{\exp\left(  iz\alpha\right)  \exp\left(
-i\pi z\right)  }{\sin\left(  \pi z\right)  }. \label{ga}%
\end{equation}
We note that in terms of $g\left(  x,y,\alpha\right)  $, $f(z)$ can be written
as
\begin{equation}
f(z)=\frac{g\left(  x,y,\alpha\right)  }{\left(  n-\beta\right)  ^{2}%
+\gamma^{2}}.
\end{equation}
With the substitution of $z=x+iy$ in Eq. (\ref{ga}), we obtain%
\begin{equation}
\left\vert g\left(  x,y,\alpha\right)  \right\vert =\frac{2\exp\left(
-y\alpha\right)  }{\left[  \exp\left(  -4\pi y\right)  -2\cos\left(  2\pi
x\right)  \exp\left(  -2\pi y\right)  +1\right]  ^{1/2}}.
\end{equation}
We note that
\begin{equation}
\lim_{\left\vert y\right\vert \rightarrow\infty}\left\vert g\left(
x,y,\alpha\right)  \right\vert =0. \label{vx}%
\end{equation}
The condition in Eq. (\ref{vx}) ensures that $I^{\prime}$ goes to zero on
those portions of the contour which lie parallel to the $x$ axis. To consider
the portions of contour parallel to the $y$ axis, we substitute $x=N+1/2$. One
obtains%
\begin{equation}
\left\vert g\left(  N+1/2,y,\alpha\right)  \right\vert =\frac{2\exp\left(
-y\alpha\right)  }{1+\exp\left(  -2\pi y\right)  },
\end{equation}
which implies that the maximum value of the function $\left\vert g\left(
N+1/2,y,\alpha\right)  \right\vert $ occurs for
\begin{equation}
y=\frac{1}{2\pi}\ln\left(  \frac{2\pi-\alpha}{\alpha}\right)  ,
\end{equation}
and is given by
\begin{equation}
g_{\text{max}}\left(  \alpha\right)  =\frac{2\pi-\alpha}{2\pi}\exp\left[
-\frac{\alpha}{2\pi}\ln\left(  \frac{2\pi-\alpha}{\alpha}\right)  \right]  .
\end{equation}
A plot of $g_{\text{max}}\left(  \alpha\right)  $ is shown in Fig.3.

\begin{figure}[ptb]
\begin{center}
\includegraphics[angle=0,scale=0.6]{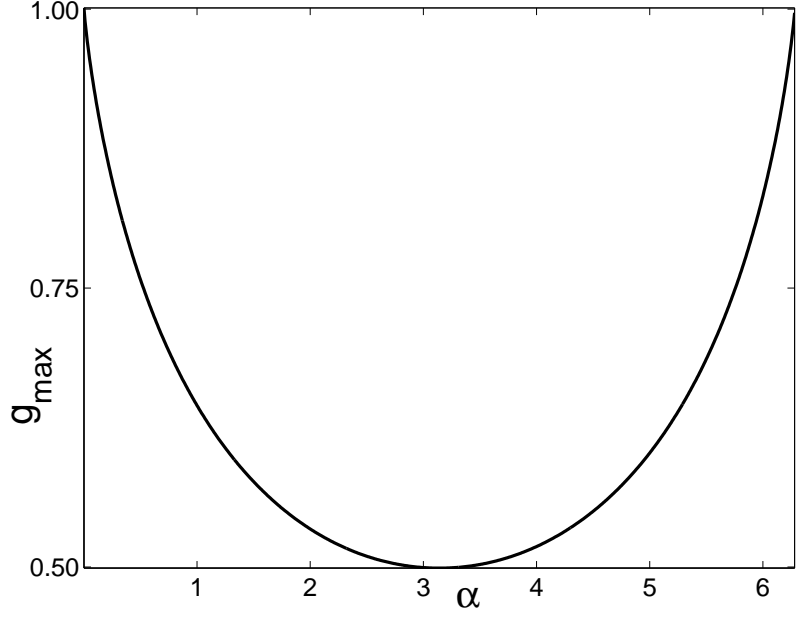}\label{gmax}
\end{center}
\caption{The value of function  $g_{\text{max}}$ remains between $0.5$
  and $1.0$ as $\alpha$ is varied from $0$ to $2\pi$. }%
\end{figure}

 It is
clear that the value of $g_{\text{max}}\left(  \alpha\right)  $ remains
between $0.5$ and $1.0$, and this ensures $|f(z)z|$ goes to zero on the
contours parallel to the $y$ axis. Thus it is clear that $I^{\prime}$ goes to
zero when evaluated for the contour $C_{N}$ and hence by the application of
formula in Eq. (\ref{a2}) we obtain%
\begin{align}
\sum_{n=-\infty}^{\infty}\frac{\exp\left(  in\alpha\right)  }{\left(
n-\beta\right)  ^{2}+\gamma^{2}}  &  =-\frac{\exp\left[  (\beta+i\gamma
)\left(  \alpha+i\pi\right)  \right]  }{2i\gamma}\pi\csc\left[  \pi
(\beta+i\gamma)\right] \\
&  -\frac{\exp\left[  (\beta-i\gamma)\left(  \alpha+i\pi\right)  \right]
}{-2i\gamma}\pi\csc\left[  \pi(\beta-i\gamma)\right] \nonumber\\
&  =\frac{\pi}{\gamma}\frac{\exp\left[  i\beta\left(  \alpha-2\pi\right)
\right]  \sinh\left(  \gamma\alpha\right)  +\exp\left(  i\beta\alpha\right)
\sinh\left[  \gamma\left(  2\pi-\alpha\right)  \right]  }{\cosh\left(
2\pi\gamma\right)  -\cos\left(  2\pi\beta\right)  }.\nonumber
\end{align}

\section{Triclinic cell}

Let us consider the most general type of triclinic cell shown in the Fig.4.
The cell is characterized by sides $l_{1}$, $l_{2}$ and $l_{3}$ and angles
$\alpha$, $\beta$ and $\gamma$. We can choose the unit vectors along the
directions of the triclinic cell as%

\begin{align}
e_{1}  &  =\left(  1,0,0\right)  ,\\
e_{2}  &  =\left(  \cos\alpha,\sin\alpha,0\right) \nonumber\\
e_{3}  &  =\left(  \cos\gamma,\frac{\cos\beta-\cos\alpha\cos\gamma}{\sin
\alpha},\left[  \sin^{2}\gamma-\left(  \frac{\cos\beta-\cos\alpha\cos\gamma
}{\sin\alpha}\right)  ^{2}\right]  ^{1/2}\right)  .\nonumber
\end{align}
We can now get reciprocal vectors using equation. Now we can calculate
$c_{ij}$ and we get the following results using the package Mathematica%

\begin{align}
c_{11}  &  =\frac{l_{3}^{2}}{l_{1}^{2}}\frac{\sin^{2}\beta}{\sin^{2}\alpha
},~c_{22}=\frac{l_{3}^{2}}{l_{2}^{2}}\frac{\sin^{2}\gamma}{\sin^{2}\alpha
},~c_{33}=1,\\
c_{12}  &  =\frac{l_{3}^{2}}{l_{1}l_{2}}\left(  \frac{-\cos\alpha+\cos
\beta\cos\gamma}{\sin^{2}\alpha}\right)  ,\nonumber\\
c_{23}  &  =\frac{l_{3}}{l_{2}}\left(  \frac{-\cos\beta+\cos\alpha\cos\gamma
}{\sin^{2}\alpha}\right)  ,\nonumber\\
c_{13}  &  =\frac{l_{3}}{l_{1}}\left(  \frac{-\cos\gamma+\cos\alpha\cos\beta
}{\sin^{2}\alpha}\right)  .\nonumber
\end{align}
We can then obtain $\rho$ and $x$ as follows%

\begin{equation}
\rho=\frac{\left(  r_{1}^{2}\cos^{2}\alpha+r_{3}^{2}\cos^{2}\beta+2r_{1}%
r_{3}\times\left[  \cos\gamma-\cos\alpha\cos\beta\right]  \right)  ^{1/2}%
}{l_{2}}%
\end{equation}
and%

\begin{figure}
\begin{center}
\includegraphics[angle=-90,scale=0.6]{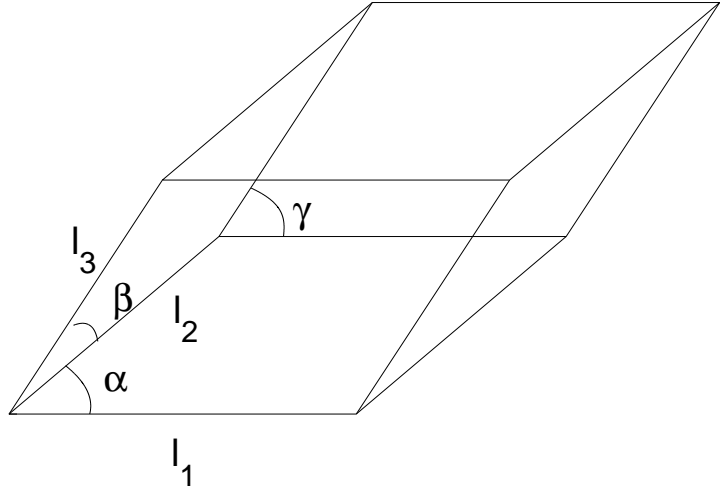}
\end{center}
\caption{A triclinic cell explaing different labels for sides and angles.}%
\end{figure}

\begin{equation}
x=\frac{r_{2}+r_{1}\cos\alpha+r_{3}\cos\beta}{l_{2}}.
\end{equation}
Finally we obtain%

\begin{equation}
\rho^{2}+x^{2}=\frac{r_{1}^{2}+r_{2}^{2}+r_{3}^{2}+2r_{1}r_{2}\cos
\alpha+2r_{2}r_{3}\cos\beta+2r_{3}r_{1}\cos\gamma}{l_{2}^{2}}.
\end{equation}
Using the relations given above, it can be shown on Mathematica that
\begin{equation}
Vb_{3}^{2}\left\vert \widetilde{\delta}\right\vert =4\pi^2 l_{2},
\end{equation}
where $b_{3}$, $V$ and $\widetilde{\delta}$ are defined in Eqs. (\ref{h10}),
(\ref{hh0}) and (\ref{h26}).

\end{document}